\documentstyle[12pt,a4wide,epsfig]{article}
\input epsf
\renewcommand{\theequation}{\arabic{section}.\arabic{equation}}
\newcommand{\news}{\setcounter{equation}{0}}
\newcommand{\be}{\begin{equation}}
\newcommand{\ee}{\end{equation}}
\newcommand{\bea}{\begin{eqnarray}}
\newcommand{\eea}{\end{eqnarray}}
\newcommand{\bean}{\begin{eqnarray*}}
\newcommand{\eean}{\end{eqnarray*}}
\font\upright=cmu10 scaled\magstep1
\font\sans=cmss12
\newcommand{\ssf}{\sans}
\newcommand{\stroke}{\vrule height8pt width0.4pt depth-0.1pt}
\newcommand{\Z}{\hbox{\upright\rlap{\ssf Z}\kern 2.7pt {\ssf Z}}}

\newcommand{\C}{{\rlap{\rlap{C}\kern 3.8pt\stroke}\phantom{C}}}
\newcommand{\R}{\hbox{\upright\rlap{I}\kern 1.7pt R}}
\newcommand{\CP}{\C{\upright\rlap{I}\kern 1.5pt P}}
\newcommand{\half}{\frac{1}{2}}
\newcommand{\identity}{{\upright\rlap{1}\kern 2.0pt 1}}
\newcommand{\bm}{\boldmath}
\newcommand{\sk}{Skyrmion}
\newcommand{\Sk}{Skyrme}
\newcommand{\I}{{\cal I}} 
\begin{document}
\pagestyle{plain}
\title{\vskip -70pt
\begin{flushright}
{\normalsize UKC/IMS/97-25,\ DAMTP 97-48,\ NI 97026-NQF,\ hep-th/9705151} \\
\end{flushright}
\vskip 20pt
{\bf \Large \bf Rational Maps, Monopoles and Skyrmions}
 \vskip 10pt
}
\author{Conor J. Houghton$^{\dagger}$,
Nicholas S. Manton$^{\dagger}$ 
and Paul M. Sutcliffe$^{\ddagger}$\\[10pt]
{\normalsize
$\dagger$ {\sl Department of Applied Mathematics 
and Theoretical Physics}}\\
{\normalsize {\sl University of Cambridge, Silver St.,
 Cambridge CB3 9EW, England.}}\\
{\normalsize {\sl Email C.J.Houghton@damtp.cam.ac.uk}}\\
{\normalsize {\sl Email N.S.Manton@damtp.cam.ac.uk}}\\[10pt]
{\normalsize 
$\ddagger$ {\sl
Institute of Mathematics,
University of Kent at Canterbury,}}\\
{\normalsize {\sl Canterbury CT2 7NF, England.}}\\
{\normalsize {\sl Email P.M.Sutcliffe@ukc.ac.uk}}\\[10pt]
}

\date{November 1997}
\maketitle

\begin{abstract}
We discuss the similarities between BPS monopoles and
Skyrmions, and point to an underlying connection in terms
of rational maps between Riemann spheres. 
This involves the introduction of a new ansatz for Skyrme
fields. We use this to construct good approximations
to several known Skyrmions, including all the minimal energy
configurations up to baryon number nine, and some new solutions
such as a baryon number seventeen Skyrme field with the
 truncated icosahedron structure of a buckyball. 

The new approach is also used to understand the low-lying
vibrational modes of Skyrmions, which are required for
quantization. Along the way we discover an interesting
Morse function on the space of rational maps which may be of
 use in understanding the Sen forms on the monopole moduli spaces.
\end{abstract}
\newpage
\section{Introduction}
\news

There is considerable evidence that there is rather a close connection
between $SU(2)$ BPS monopoles and \sk s, despite their obvious differences.

Recall that BPS monopoles are minimal energy classical solutions in a
Yang-Mills-Higgs theory. They satisfy the Bogomolny equation
\be
B_i=D_i\Phi
\label{bog}
\ee
where $B_i$ is the magnetic part of the $SU(2)$ Yang-Mills field
tensor, and $D_i\Phi$ is the covariant derivative of an adjoint Higgs
field. There is a $(4N-1)$-dimensional moduli space of gauge
inequivalent solutions with monopole number $N$, all with the same
energy $4\pi N$. Among these solutions are some special ones of rather
high symmetry, representing $N$ coalesced single monopoles.

The Skyrme model is a nonlinear theory of pions, with an $SU(2)$ valued scalar field $U({\bf x},t)$,
the Skyrme field, satisfying the boundary condition
$U\rightarrow 1$ as ${ \vert {\bf x}\vert \rightarrow\infty}.$ Static
fields obey the equation
\be
\partial_i(R_i-\frac{1}{4}[R_j,[R_j,R_i]])=0
\label{sky}
\ee
where $R_i$ is the $su(2)$ valued current $R_i=(\partial_iU)U^{-1}.$
Such fields are stationary points (either minima or saddle points) of the energy
function
\be
E=\int \{-{1 \over 2}\mbox{Tr}(R_iR_i)-{1 \over 16}
\mbox{Tr}([R_i,R_j][R_i,R_j])\} \ d^3{\bf x}.
\label{energy}
\ee

Associated with a Skyrme field is a topological integer, the baryon
number $B$, defined as the degree of the map $U: \R^3\mapsto SU(2)$.
This is calculated at a given time, but is time independent. It is
well defined because of the boundary condition at infinity.

Solutions of the Skyrme equation (\ref{sky}) are known for several values of $B$,
but they can only be obtained numerically. Many of these solutions are
stable, and probably represent the global minimum of the energy for
given $B$. We shall refer to the solutions believed to be of lowest
energy for each $B$ as \sk s. Some saddle-point solutions are also
known.

All known solutions appear to be isolated and their only moduli are
the obvious ones associated with the large symmetry group of the
equation and boundary condition. This symmetry group is
nine-dimensional. It consists of translations and rotations in $\R^3$
and the $SO(3)$ isospin transformations $U\mapsto {\cal O}U{\cal
  O}^{-1}$ where ${\cal O}$ is a constant element of $SU(2)$. Generic
solutions therefore have nine moduli, although solutions with axial or
spherical symmetry have fewer.

It has been found that many solutions of the Skyrme equation,
and particularly those of low energy, look rather like
monopoles, with the baryon number $B$ being identified with the
monopole number $N.$ The fields are not really the same, but the
energy density has equivalent symmetries and approximately the same
spatial distribution. The history of the discovery of these solutions
is perhaps worth recalling. First Skyrme \cite{Sk} found the
spherically symmetric $B=1$ \sk.  Later Prasad \& Sommerfield
\cite{PS} found the analytic form of the $N=1$ monopole, which is also
spherically symmetric. Bogomolny \cite{Bo} then showed that minimal
energy monopoles should satisfy (\ref{bog}), and that the
Prasad-Sommerfield solution is the unique spherically symmetric
solution of (\ref{bog}). Next, and with difficulty, an axially
symmetric $N=2$ monopole was discovered by Ward \cite{Wa1}, and after
that it was found that the $B=2$ \sk\ is axially
symmetric \cite{KS,Ma3,V}.  (There are spherically symmetric solutions
of (\ref{sky}) for all $B$, but for $B>1$ they have rather high
energy.)

Next, a substantial numerical search for \sk\ solutions was undertaken
by Braaten et al. \cite{BTC}, and minimal energy solutions up to $B=6$
were constructed.  (Their solution for $B=6$ was rather inaccurate,
and its symmetry was wrongly identified.) Surprisingly, the $B=3$
solution has tetrahedral symmetry $T_d$, and the $B=4$ solution has
octahedral symmetry $O_h.$ The $B=5$ (and $B=6$) solutions have lower
symmetry.  These findings suggested that monopoles with similar
symmetries might exist, and indeed they do.

Hitchin et al. \cite{HMM} established the existence of an $N=3$
monopole with tetrahedral symmetry and an $N=4$ monopole with octahedral
symmetry. These solutions are unique, up to the action of the
Euclidean group. All other BPS monopole solutions with these monopole
numbers (other than the axially symmetric configurations) have lower
symmetry. Hitchin et al. also showed that no monopoles with
icosahedral symmetry are possible with $N\le 6$ (although it appeared
at first that one with $N=6$ might be possible). Subsequently,
Houghton \& Sutcliffe \cite{HS2} found an $N=5$ monopole with octahedral
symmetry, and an $N=7$ monopole with icosahedral symmetry.

The search was on for further solutions of the Skyrme equation. There
is a $B=5$ solution with octahedral symmetry, but it has slightly higher
energy than the minimal energy solution \cite{BS3}. More importantly,
Battye \& Sutcliffe \cite{BS2} established that the $B=7$ \sk\ has
icosahedral symmetry. Battye \& Sutcliffe \cite{BS2} have recently
found all \sk s up to $B=9.$ The $B=6$ and $B=8$ solutions have the
relatively low symmetries $D_{4d}$ and $D_{6d}$ respectively, but the
$B=9$ solution has tetrahedral symmetry. The results in this paper
imply that monopoles with these symmetries, for $N=6,8$ and $9$
respectively, exist too, but little is known about them.

Pictures of all these Skyrmion solutions can be found in ref.
\cite{BS2}. Qualitatively, they are like the pictures in Fig. 1, whose
significance we will explain later.

As we mentioned earlier, Skyrmion solutions are isolated, but it is
physically interesting to study the small oscillation vibrations around
them.  The vibrational modes of the axisymmetric $B=2$ Skyrmion and
the octahedrally symmetric $B=4$ Skyrmion have recently been studied
by Barnes et al. \cite{BBT1,BBT2}. The frequencies and degeneracies of
these modes have been calculated, and the way the Skyrmion vibrates
can be visualized. At least for these two examples, the lowest
frequency modes can clearly be identified with the deformations of the
moduli of the corresponding symmetric monopole. For example, 
for the octahedrally symmetric $N=4$ monopole
there are fifteen moduli in all, but six of these are associated with
the Euclidean group. The nine remaining moduli can be identified with
the nine lowest frequency vibrational modes of the $B=4$ \sk.

How can we understand all these results? The aim of this paper is to
point to an explanation in terms of rational maps. A
rational map is a holomorphic function from $S^2\mapsto S^2$. If we
treat each $S^2$ as a Riemann sphere, the first having coordinate $z$,
a rational map of degree $N$ is a function $R:S^2\mapsto S^2$ where
\be 
R(z)=\frac{p(z)}{q(z)} \label{rat}
\ee 
and $p$ and $q$ are polynomials of degree at most $N$. At least
one of $p$ and $q$ must have degree precisely $N$, and $p$ and $q$
must have no common factors (ie. no common roots).

Rational maps were introduced into the theory of monopoles by
Donaldson \cite{Do}. Indeed Donaldson showed that there is a
one-to-one correspondence between maps of degree $N$ (with the basing
condition $R(z)\rightarrow 0$ as $z\rightarrow\infty$) and $N$-monopoles.  
Donaldson's work, following Hitchin \cite{Hi1,Hi2}, relies
on a choice of direction in $\R^3$, and this is not helpful in the
present context.

A new relationship between monopoles and rational maps has recently
been established by Jarvis \cite{Ja} (following a suggestion of
Atiyah). This requires the choice of an origin, and is much better
adapted for studying fields invariant under a subgroup of the group of
rotations about the origin. The Jarvis map is obtained by considering
Hitchin's equation 
\be (D_r-i\Phi)s=0
\label{hitchin}
\ee along each radial line from the origin to infinity. Here $D_r$ is
the covariant derivative in the radial direction and $\Phi$ is the
Higgs field. $s$ is an auxiliary complex doublet field transforming
via the fundamental representation of the gauge group $SU(2)$.
Because $\Phi$ is asymptotically conjugate to
$\mbox{diag}(\frac{i}{2},-\frac{i}{2})$, equation (\ref{hitchin}) has,
up to a constant multiple, just one solution which decays
asymptotically as $r\rightarrow\infty.$ Let ${\scriptsize
  \bigg(\begin{array}{c}s_1(r)\\s_2(r)\end{array}\bigg)}$ be this
solution and ${\scriptsize \bigg(\begin{array}{c}s_1(0)\\s_2(0)\end{array}\bigg)}$
its value at the origin. Because of the arbitrariness of the constant
multiple, it is only the ratio $R=s_1(0)/s_2(0)$ that is interesting.
Now a particular radial line is labelled by its direction, regarded as
a point $z$ on the Riemann sphere. $R$ depends holomorphically on the
direction $z$, so we write $R(z)$. The reason $R$ is holomorphic is
that the complex covariant derivative in the angular direction,
$D_{\bar z}$, commutes with the operator $D_r-i\Phi$, because of the
Bogomolny equation (\ref{bog}).  It can be shown that the degree of
$R$ is equal to the monopole number $N$, and hence $R$ is rational of
degree $N$. There is one remaining ambiguity in $R(z)$. If we carry
out a gauge transformation then $R(z)$ is replaced by its M\"obius
transformation by an $SU(2)$ matrix \be R(z)\mapsto \frac{\alpha
  R(z)+\beta}{-\bar\beta R(z)+\bar\alpha} 
\label{mobius}
\ee 
with $\vert\alpha\vert^2+\vert\beta\vert^2=1$. The $SU(2)$ matrix here, ${\scriptsize \bigg(\begin{array}{cc}\alpha & \beta\\
    -\bar\beta & \bar\alpha\end{array}\bigg)}$, is the gauge
transformation matrix evaluated at the origin, and it acts globally on
$R(z)$, that is, the same matrix occurs for all $z$.

Thus the moduli space of rational maps $R(z)$ that Jarvis associates
with $N$-monopoles is the complete $(4N+2)$-dimensional space of
unbased rational maps of degree $N$. For each monopole there is
a map which is uniquely defined up to an $SU(2)$ M\"obius
transformation.  Moreover, each rational map arises from some
monopole. Jarvis shows how, in principle, one may directly reconstruct
the monopole from the rational map. This algorithm could be
implemented numerically, and currently work is in progress to achieve
this. Note that this construction still requires the solution of a
partial differential equation in three-dimensional space so the
computational gain is small. However, its advantage over a direct
numerical solution of the Bogomolny equation is that the selection of
a given monopole solution can be made precise via the rational map
input, which is easy to obtain. This contrasts with an existing
numerical construction \cite{HS1}, where the computational gain is
great, since only ordinary differential equations need to be solved,
but where the input is more difficult to obtain since it consists of
Nahm data which can only be found after the solution of a nonlinear
system of matrix differential equations.

The naturalness of the Jarvis construction means that a monopole invariant
under a subgroup $G$ of the spatial rotation group $SO(3)$ will have
an associated map $R(z)$ which is $G$-invariant (up to M\"obius
transformations), and conversely, if we find a $G$-invariant
map of a given degree $N$ then there is an $N$-monopole with symmetry
$G$. If we find the complete set of rational maps invariant under
$G$, then the corresponding set of $G$-invariant monopoles will
form a geodesic submanifold of the monopole moduli space. In
particular, if, for some $G$, the set of maps is one-dimensional, then
the corresponding monopoles lie on a geodesic in the moduli space.
Using the geodesic approximation to monopole motion \cite{Ma1}, we
obtain, usually, an example of monopole
scattering with $G$-invariance.

An important quantity associated with a rational map $R(z)=p(z)/q(z)$ is the Wronskian
\be
W(z)=p'(z)q(z)-q'(z)p(z)
\label{wron}
\ee or more precisely, the zeros of $W$, which are the branch
points of the map.  If $R$ is of degree $N$, then generically, $W$
is a polynomial of degree $2N-2.$ The zeros of $W$ are invariant
under any M\"obius transformation of $R$, which replaces $p$ by
$\alpha p+\beta q$ and $q$ by $\gamma q+\delta p$ and hence simply
multiplies $W$ by $(\alpha\gamma-\beta\delta).$ Occasionally,
$W$ is a polynomial of degree less than $2N-2$, but one then
interprets the missing zeros as being at $z=\infty.$ The symmetries of
the map $R$, and hence of the $N$-monopole which corresponds to it,
are captured by the symmetries of the Wronskian $W$. Sometimes $W$
has more symmetry than the rational map $R$, and we shall see
examples of this.

Monopoles with given symmetries have been constructed before, for
example, the $N=7$ monopole with icosahedral symmetry $Y_h.$ But the
construction depended on a careful study of Nahm's equation, and the
existence of the solution was not known in advance. The Nahm equation
approach has only been successfully applied in relatively simple
cases. It is much easier to classify rational maps with given
symmetries, and we shall give a number of examples later. This
establishes the existence of monopoles with these symmetries,
but we have not constructed all the solutions, even numerically!

\section{Skyrme Fields from Rational Maps}
\news

The understanding of monopoles in terms of rational maps suggests that
one might understand a range of \sk\ solutions using rational maps.
Rational maps are maps from $S^2\mapsto S^2$, whereas \sk s are maps
from $\R^3\mapsto S^3$. A rather naive idea, which we find works quite
well, is to identify the domain $S^2$ with concentric spheres in
$\R^3$, and the target $S^2$ with spheres of latitude on $S^3.$ This
leads to a new ansatz for Skyrme fields.

It is convenient to use Cartesian notation to present the ansatz.
Recall that via stereographic projection, the complex coordinate $z$
on a sphere can be identified with conventional polar coordinates by
$z=\tan(\theta/2)e^{i\varphi}.$ Equivalently, the point $z$
corresponds to the unit vector 
\be 
\widehat{\bf n}_z=\frac{1}{1+\vert
  z\vert^2}(2\Re(z), 2\Im(z),1-\vert z\vert^2).
\label{unit1}
\ee
Similarly the value of the rational map $R(z)$ is associated with the unit vector
\be
\widehat{\bf n}_R=\frac{1}{1+\vert R\vert^2}(2\Re(R), 2\Im(R),1-\vert R\vert^2).
\label{unit2}
\ee

Let us denote a point in $\R^3$ by its coordinates $(r,z)$ where $r$
is the radial distance from the origin and $z$ specifies the direction
from the origin. Our ansatz for the Skyrme field depends on a rational
map $R(z)$ and a radial profile function $f(r).$ The ansatz is \be
U(r,z)=\exp(if(r) \ \widehat{\bf n}_R\cdot\mbox{\bm $\sigma$})
\label{ansatz}
\ee
where $\mbox{\bm $ \sigma$}=(\sigma_1,\sigma_2,\sigma_3)$ denotes the Pauli matrices.
For this to be well-defined at the origin, $f(0)=k\pi$, for some integer $k.$
The boundary value $U=1$ at $r=\infty$ requires that $f(\infty)=0.$ It is straightforward
to verify that the baryon number of this field is $B=Nk$, where $N$ is the degree of $R.$
In the remainder of this paper we shall only consider the case $k=1$,
and then $B=N.$ Note that an $SU(2)$ M\"obius transformation on the
target $S^2$ of the rational map corresponds to a rotation of 
$\widehat{\bf n}_R$, and hence to an isospin rotation of the Skyrme field.

In the case $N=1$, the basic map is $R(z)=z$, and (\ref{ansatz}) reduces to
Skyrme's hedgehog field
\be
U(r,\theta,\varphi)=\cos f+i\sin f(\sin\theta\cos\varphi\ \sigma_1
+\sin\theta\sin\varphi\ \sigma_2+\cos\theta\ \sigma_3).
\label{hh}
\ee
The simplest case beyond this, with $N=2$, is $R=z^2$, which gives an
ansatz rather different from that tried in ref. \cite{WSG} for the
$B=2$ \sk. We shall return to this case in more detail in Section 3.

An attractive feature of the ansatz (\ref{ansatz}) is that it leads to
a simple energy expression which can be minimized with respect
to the rational map $R$ and the profile function $f$ to obtain close
approximations to several known \sk\ solutions. Starting with these
approximations is an efficient method to find new exact solutions,
although we shall not pursue this application here. To calculate the
energy of a field of the form (\ref{ansatz}) we exploit an
interpretation of the Skyrme energy function given in ref. \cite{Ma2}.

As in nonlinear elasticity theory, the energy density of a Skyrme
field depends on the local stretching associated with the map
$U:\R^3\mapsto S^3.$ The Riemannian geometry of $\R^3$ (flat) and of
$S^3$ (a unit radius 3-sphere) are necessary to define this
stretching.  Consider the strain tensor at a point in $\R^3$ \be
D_{ij}=-{1 \over 2}\mbox{Tr}(R_iR_j)=-{1 \over 2}\mbox{Tr}
((\partial_iUU^{-1})(\partial_jUU^{-1})).
\label{strain}
\ee
This is symmetric, and positive semi-definite as $R_i$ is
antihermitian. Let its eigenvalues be $\lambda_1^2$,
$\lambda_2^2$ and $\lambda_3^2.$ The Skyrme energy can be reexpressed as
\be
E=\int 
(\lambda_1^2 +\lambda_2^2 +\lambda_3^2+ \lambda_1^2\lambda_2^2+ \lambda_2^2\lambda_3^2+
\lambda_1^2\lambda_3^2)
 \ d^3{\bf x},
\label{energy2}
\ee
and the baryon density as $\lambda_1\lambda_2\lambda_3 / 2\pi^2$.
For the ansatz (\ref{ansatz}), the strain in the radial direction is orthogonal to the strain
in the angular directions. Moreover, because $R(z)$ is conformal, the angular strains are
isotropic. If we identify $\lambda_1^2$ with the radial strain and $\lambda_2^2$ and
$\lambda_3^2$ with the angular strains, we can easily compute that
\be
\lambda_1=-f'(r), \hskip 1cm 
\lambda_2=\lambda_3=\frac{\sin f}{r}\frac{1+\vert z\vert^2}{1+\vert R\vert^2}
\bigg\vert\frac{dR}{dz}\bigg\vert.
\label{strain2}
\ee
Therefore the energy is
\bea
E=\int\bigg[
f'^2+2(f'^2&+&1)\frac{\sin^2f}{r^2}
\bigg(
\frac{1+\vert z\vert^2}{1+\vert R\vert^2}
\bigg\vert\frac{dR}{dz}\bigg\vert\bigg)^2\label{energy3}\\
&+&
\frac{\sin^4f}{r^4}
\bigg(
\frac{1+\vert z\vert^2}{1+\vert R\vert^2}
\bigg\vert\frac{dR}{dz}\bigg\vert\bigg)^4\bigg]
\frac{2i \  dzd\bar z r^2  dr}{(1+\vert z\vert^2)^2}, 
\nonumber
\eea
where $2i \   dz  d\bar z  /(1+\vert z\vert^2)^2$
is equivalent to the usual area element on a 2-sphere $\sin\theta d\theta  d\varphi.$
Now the part of the integrand
\be
\bigg(
\frac{1+\vert z\vert^2}{1+\vert R\vert^2}
\bigg\vert\frac{dR}{dz}\bigg\vert\bigg)^2
 \frac{2i \   dz  d\bar z }{(1+\vert z\vert^2)^2}
\label{winding}
\ee
is precisely the pull-back of the area form $2i \  dRd\bar R/(1+\vert R\vert^2)^2$
on the target sphere of the rational map $R$; therefore its integral is $4\pi$ times
the degree $N$ of $R$. So the energy simplifies to
\be
E=4\pi\int \bigg(
r^2f'^2+2N(f'^2+1)\sin^2 f+\I\frac{\sin^4 f}{r^2}\bigg) \ dr
\label{energy4}
\ee
where $\I$ denotes the integral
\be
\I=\frac{1}{4\pi}\int
 \bigg(
\frac{1+\vert z\vert^2}{1+\vert R\vert^2}
\bigg\vert\frac{dR}{dz}\bigg\vert\bigg)^4 \frac{2i \  dz  d\bar z }{(1+\vert z\vert^2)^2}.
\label{i}
\ee
$\I$ depends only on the rational map $R$, and, as we explain in
Section 5, it is an interesting
function on the space of rational maps.

To minimize $E$, for maps of a given degree
$N$, one should first
minimize $\I$ over all maps of degree $N$. Then,
the profile function $f$ 
minimizing the energy (\ref{energy4}) may be found by solving a second
order differential equation with $N$ and $\I$ as parameters. 
In practice, we have considered rational maps of a given symmetric
form, with symmetries corresponding to a known \sk\ solution (or
monopole). If these maps still contain a few free parameters, we have
minimized $\I$ with respect to these (using an appropriate search
algorithm). Then, the minimizing
profile function $f$ is determined by first bijectively mapping the radial
coordinate $r$ onto the unit interval, and then discretizing the
energy functional using symmetric finite differences and finally
obtaining the minimizing grid values by applying a conjugate gradient
algorithm. 
This procedure seems appropriate for
all baryon numbers up to $B=9$, where the \sk\ solutions all have
considerable symmetry, but for some higher values of $B$ one will
have to consider quite general maps as the Skyrmions
probably have very little symmetry.

Detailed examples of rational maps with various degrees and symmetries
will be described in the next Section, and we shall compare the result
of minimizing $E$ for these maps with the energies of the numerically
determined exact \sk\ solutions.

Note the following pair of inequalities associated with the expression
(\ref{energy4}) for the energy $E$. The elementary inequality 
\be
\bigg(\int 1\ dS\bigg) \bigg(\int \bigg(\frac{1+\vert
  z\vert^2}{1+\vert R\vert^2}
\bigg\vert\frac{dR}{dz}\bigg\vert\bigg)^4 \ dS\bigg) \ge \bigg(\int
\bigg(\frac{1+\vert z\vert^2}{1+\vert R\vert^2}
\bigg\vert\frac{dR}{dz}\bigg\vert\bigg)^2 \ dS\bigg)^2,
\label{ineq1}
\ee
where $dS=2i \  dz  d\bar z  /(1+\vert z\vert^2)^2$, implies that $\I\ge N^2.$ Next, by a
Bogomolny-type argument, we see that
\be
E=4\pi\int \left(
(rf'+\sqrt{\I}\frac{\sin^2 f}{r})^2+2N(f'+1)^2\sin^2 f-2(2N+\sqrt{\I})f'\sin^2 f\right) dr
\label{energy5}
\ee
so
\be
E\ge 4\pi(2N+\sqrt{\I})\int_0^\infty (-2f'\sin^2 f) \ dr
=4\pi(2N+\sqrt{\I})\bigg[-f+\half\sin 2f\bigg]_0^\infty
\label{energy6}
\ee
and so, if $f(0)=\pi$ and $f(\infty)=0$,
\be
E\ge 4\pi^2(2N+\sqrt{\I}).
\label{bound}
\ee Combined with the earlier inequality for $\I$, we recover the
usual Fadeev-Bogomolny bound $E\ge 12\pi^2N.$ The bound (\ref{bound})
is stronger than this, for fields of the form we are considering, but
there is no reason to think that true solutions of the Skyrme
equation are constrained by this bound.

We conclude this Section by observing that the zeros of the Wronskian
$W(z)$ of a rational map $R(z)$ give interesting information about the
shape of the Skyrme field which is constructed from $R$ using our
ansatz (\ref{ansatz}). Where $W$ is zero, the derivative $dR/dz$ is
zero, so the strain eigenvalues in the angular directions, $\lambda_2$
and $\lambda_3$, vanish. The baryon density, being proportional to
$\lambda_1\lambda_2\lambda_3$, vanishes along the entire radial line
in the direction specified by any zero of $W$. The energy density
will also be low along such a radial line, since there will only be the
contribution $\lambda_1^2$ from the radial strain eigenvalue. The Skyrme
field baryon density contours will therefore look like a polyhedron with holes
in the directions given by the zeros of $W$, and there will be
$2N-2$ such holes. This structure is seen in all the plots shown in
Fig. 1, for example, the $B=7$ \sk\ having twelve holes arranged at the
face centres of a dodecahedron.

\section{Symmetric Rational Maps and Skyrmions}
\news

In this Section, we present the detailed form of certain symmetric
rational maps of degrees one to nine, and also of degrees eleven and
seventeen. Using our ansatz (\ref{ansatz}) we turn these rational maps
into Skyrme fields with baryon number equal to the degree of the map.
In each case, except degree eleven, we determine the parameters of the
rational map that
minimize its contribution to the energy, and then find the profile
function $f(r)$ which minimizes the energy function (\ref{energy4})
using the method explained earlier.  In Fig. 2 we plot these profile
functions for baryon numbers one to nine and also seventeen. The size
increases with increasing baryon number, corresponding to a shift to
the right of the profile function, hence they need not be labelled
individually. In Table 1 we present the values of the energy of the
resulting Skyrme field; these can be compared with the bound
(\ref{bound}), and also with the energy of the corresponding
numerically known exact Skyrmion solution.  All
numerical values for the energies quoted in this Section are the real
energies divided by $12\pi^2B$, and hence close to unity. In Fig.  1
we plot a surface of constant baryon density for several of our
computed \Sk\ fields. It should be noted that the value of the baryon
density on the surface shown is not the same in each case, so the
scale of these pictures should not be used to infer information on the
size.  However, the size can be deduced from the profile plots in Fig.
2. The data on the energies and shapes show that the Skyrme fields we
obtain closely approximate the true Skyrmions.

The symmetries we impose are not chosen systematically; they are
motivated by the symmetries of the known \sk\ solutions
(for $B\le 9$) \cite{BS2}. For $B=11$ and $B=17$ we impose icosahedral
symmetry, which is a possibility in both cases, and which has been
conjectured as the symmetry of the $B=17$ Skyrmion. 

Although most of the approximate solutions we find using our ansatz
are minima of the energy, some are saddle points. True saddle point
solutions of the Skyrme equation are often the minimal energy
configurations having a particular symmetry not possessed by the
Skyrmion solution.

A rational map, $R:S^2\mapsto S^2$, is invariant or symmetric under a
subgroup $G\subset SO(3)$ if there is a set of M\"obius transformation
pairs $\{g,D_g\}$ with $g\in G$ acting on the domain $S^2$ and $D_g$
acting on the target $S^2$, such that \be R(g(z))=D_gR(z).
\label{inv}
\ee
The transformations $D_g$ should represent 
 $G$ in the sense that
\hbox{$D_{g_1}D_{g_2}=D_{g_1g_2}.$}
Both $g$ and $D_g$ will in practice be $SU(2)$ matrices. For example, $g(z)$ can be expressed
as $g(z)=(\alpha z+\beta)/(-\bar\beta z+\bar\alpha)$ with 
$\vert\alpha\vert^2+\vert\beta\vert^2=1.$ Replacing $(\alpha,\beta)$ by $(-\alpha,-\beta)$
has no effect, so $g$ is effectively in $SO(3)$. The same is true for $D_g.$

Some of our rational maps possess an additional reflection or
inversion symmetry. The transformation $z\mapsto \bar z$ is a
reflection, whereas $z\mapsto -1/\bar z$ is the antipodal map on
$S^2$, or inversion. We shall deal with reflection and inversion on a
case by case basis.

The detailed form of our maps will depend on choices of the
orientation of axes, both in the domain $S^2$ and target $S^2.$ Our
choice is made to simplify our maps as far as possible, but equivalent
maps, differently oriented, are sometimes advantageous.

It is helpful to identify the Cartesian axes with certain directions
specified by values of $z$. The formula (\ref{unit1}) for a unit
vector associated with $z$ implies that the positive $x_3$-axis is in
the direction $z=0$, the positive $x_1$-axis corresponds to $z=1$ and
the
positive $x_2$-axis corresponds to $z=i.$\\

\subsection*{$N=1$}

The hedgehog map is $R(z)=z$. It is fully $O(3)$ invariant, since
$R(g(z))=g(z)$ for any $g \in SU(2)$ and $R(-1/\bar z)=-1/\bar R(z).$
It gives the standard exact hedgehog Skyrmion solution with the usual
profile $f(r)$, and with energy $E=1.232.$ The map $R(z)=z$ is also
the Jarvis rational map of a monopole centred at the origin.\\

\subsection*{$N=2$}

A general degree two map is of the form 
\be
R(z)=\frac{\alpha z^2+\beta z+\gamma }{\lambda z^2+\mu z+\nu}. 
\ee 
Let us impose the two $\Z_2$
symmetries $z\mapsto -z$ and $z\mapsto 1/z$ which generate the
viergruppe of $180^\circ$ rotations about all three Cartesian axes.
The conditions \be R(-z)=R(z) \hskip 1cm \mbox{and} \hskip 1cm
R(1/z)=1/R(z)
\label{z2}
\ee
restrict $R$ to the form
\be
R(z)=\frac{z^2-a}{-az^2+1}.
\label{g2}
\ee
By a target space M\"obius transformation, we can bring $a$ to lie in
the interval $-1 \le a \le 1$, with the map degenerating at the
endpoints. Further, a $90^\circ$ rotation, $z \mapsto iz$, reverses
the sign of $a$. The maps (\ref{g2}) have three reflection symmetries
in the Cartesian axes, which are manifest when $a$ is real. For
example, $R(\bar z)=\bar R(z)$ when $a$ is real.

The Jarvis map of any centred and suitably oriented $N=2$ monopole is
of this form. When $a=0$ the rational map has the additional symmetry
$R(e^{i\chi}z)=e^{2i\chi}R(z)$; it is the Jarvis map of the axially
symmetric $N=2$ monopole. The maps (\ref{g2}), with $-1 < a < 1$,
parametrize a geodesic in the monopole moduli space, along which two
monopoles scatter by $90^\circ$ symmetrically from the $x_1$-axis to
the $x_2$-axis. As $a \rightarrow \pm 1$ the monopoles separate to
infinity.

If we use the maps (\ref{g2}) in our ansatz for the Skyrme field, we
find the integral $\I$ increases monotonically to infinity as $a$
increases from $0$ to $1$.  For $a=0$,
$\I=\pi+8/3$ and after determining the profile $f(r)$ in this case we
obtain $E=1.208.$ So the Skyrme field based on the map $R(z)=z^2$ has
the same symmetry as the $B=2$ \sk\ and energy just 3\% higher (see
Table 1). A baryon density plot for this configuration is shown in
Fig. 1a.

One might consider imposing $\Z_2\times\Z_2$ symmetry in other ways
than (\ref{z2}) (eg. $R(-z)=-R(z)$) but this leads to maps which are
equivalent but differently oriented.\\

\subsection*{$N=3$}

We recall that there is a unique tetrahedrally symmetric $N=3$
monopole, and that the $B=3$ \sk\ has the same symmetry. There is also
an axially symmetric, toroidal monopole (as for all $N>1$), and a
saddle point solution of the \Sk\ equations with this shape.

A subset of the degree three rational maps which allows for both these
solutions and a smooth interpolation between them is the subset with
$\Z_2\times\Z_2$ symmetry, realized by the requirements \be
R(-z)=-R(z) \hskip 1cm \mbox{and} \hskip 1cm R(1/z)=1/R(z).
\label{z2b}
\ee The first condition implies that the numerator of $R$ is even in
$z$ and the denominator is odd, or vice versa. These two
possibilities are related by an $SU(2)$ M\"obius transformation, so we
choose the former and ignore the latter. Imposing the second condition
as well gives us maps of the form \be
R(z)=\frac{\sqrt{3}az^2-1}{z(z^2-\sqrt{3}a)}
\label{g3}
\ee
with $a$ complex. The inclusion of the $\sqrt{3}$ factor is a convenience.
The parameter space of these maps should be thought of as a Riemann
sphere with complex coordinate $a.$ The rational map degenerates for
three values of $a$, namely $a=\infty$, $a=\pm 1/\sqrt{3}.$ 

There is a
further reflection symmetry $R(\bar z)=\bar R(z)$ if $a$ is real.
Together with the rotational symmetries, this implies reflection
symmetry in all three Cartesian axes. A slightly subtler symmetry
occurs if $a$ is imaginary.  The full symmetry group becomes $D_{2d}$,
where the extra generator is a rotation by $90^\circ$ about the
$x_3$-axis combined with the reflection $x_3\mapsto -x_3.$ On the
$z$-sphere the generator is $z\mapsto i/\bar z$, and $R(i/\bar
z)=i/\bar R(z)$ if $a$ is imaginary.

Tetrahedral symmetry is obtained by imposing the further symmetry
\be
R\bigg(\frac{iz+1}{-iz+1}\bigg)=\frac{iR(z)+1}{-iR(z)+1}
\label{s3}
\ee which is satisfied by (\ref{g3}) if $a=\pm i.$ Note that
\hbox{$z\mapsto (iz+1)/(-iz+1)$} sends \hbox{$0\mapsto 1\mapsto
  i\mapsto 0$} and hence generates the $120^\circ$ rotation cyclically
permuting the Cartesian axes.

Finally, there is axial symmetry about the $x_3$-axis when $a=0$,
since then $R(z)=-1/z^3.$ There is also axial symmetry when $a=\pm
\sqrt{3}.$ These further solutions are related to the first by
$120^\circ$ rotations that take the $x_3$-axis to the $x_1$-axis and
$x_2$-axis.

The $a$-sphere, with the special points we have discussed, is sketched
in Fig. 3. The Jarvis maps of the form (\ref{g3}) parametrize a
geodesic submanifold of the $N=3$ monopole moduli space.  One
particular geodesic is the great circle segment
$-1/\sqrt{3}<a<1/\sqrt{3}.$ This describes $90^\circ$ scattering of
monopoles, with two single monopoles scattering from the $x_1$-axis to
the $x_2$-axis, the third monopole remaining at the origin. This type
of geodesic was previously described by Bielawski \cite{Bi2} and
Houghton \& Sutcliffe \cite{HS5}. A second geodesic is the great
circle $-\infty<ia<\infty$, which passes through both tetrahedra and
one of the tori.  This is the twisted line scattering described in
\cite{HS3}.  Dynamical simulations of the \Sk\ equation have revealed
that remarkably similar scattering processes also occur for \sk s
\cite{BS1,BS3}.

Using (\ref{g3}) in the rational map ansatz for \Sk\ fields, we find
that on the $a$-sphere, the angular integral $\I$ has just two types
of stationary point. There are minima at the tetrahedral points $\pm
i$, where $\I=13.58$, and there are saddle points at the tori $a=0,\pm
\sqrt{3}$, where $\I=18.67.$ $\I$ diverges as the degenerate points
are approached. Using the tetrahedral map and solving for the profile
$f(r)$, we find an approximation to the tetrahedral \sk\ with energy
$E=1.184$ (see Fig. 1b).  Similarly, using the toroidal map and again
solving for the profile $f(r)$, we find an approximation to the toroidal
saddle point solution of the \Sk\ equations, with energy $E=1.256.$

In addition to the tetrahedral and toroidal solutions of the Skyrme
equation, there is a pretzel, or figure eight shaped solution,
discovered in an approximate form by Walet \cite{Wa}.  This is a
saddle point and slightly lower in energy than the torus. Its
existence has been confirmed using a full field simulation and its
energy computed to be $E=1.164$ \cite{BS3}. One might expect, based on
symmetry, that this solution could be described approximately with our
rational map ansatz. It would occur for a map of type (\ref{g3}), with
$a$ in the range $0<a<1/\sqrt{3}.$ However, no saddle point occurs in
this range. So the pretzel solution is not accessible with the
rational map ansatz, and this appears to be because it is a
configuration of three \sk s in a line, whereas the rational map ansatz
appears to work best for shell-like structures, where all the baryon
density is concentrated at roughly the same distance from the origin.

It is interesting to look at the Wronskian of maps of the form
(\ref{g3}). Recall that $W=p'q-q'p$, where $p$ and $q$ are the
numerator and denominator. Calculating, we find \be
W(z)=-\sqrt{3}a(z^4+\sqrt{3}(a-a^{-1})z^2+1).  \ee Note that for $a=\pm
i$, $W$ is proportional to a tetrahedral Klein polynomial \cite{Kl}.
If $a=1$, $W$ has square symmetry, but the rational map does
not have as much symmetry as this.\\

\subsection*{$N=4$}

The minimal energy $B=4$ \sk\ has octahedral symmetry, and there is a
unique octahedrally symmetric $N=4$ monopole. The octahedrally symmetric
rational map of degree four can be embedded in a one parameter family
of tetrahedrally symmetric maps \be
R(z)=c\frac{z^4+2\sqrt{3}iz^2+1}{z^4-2\sqrt{3}iz^2+1}
\label{g4}
\ee where $c$ is real.  The numerator and denominator are
tetrahedrally symmetric Klein polynomials, so $R$ is invariant up to a
constant factor under any transformation in the tetrahedral group.

For $c=1$ there is octahedral symmetry. The extra generator is a
$90^\circ$ rotation about the $x_3$-axis, $z\mapsto iz.$ Clearly
\be
R(iz)=1/R(z)
\label{extracubicgen}
\ee when $c=1.$ More generally, this $90^\circ$ rotation replaces $c$
by $1/c.$ There is a geodesic motion of monopoles, with tetrahedral
symmetry throughout, in which four single monopoles approach on the
vertices of a contracting tetrahedron, and recede on the vertices of
an expanding tetrahedron dual to the first \cite{HS1}. This just
corresponds to $c$ running from $0$ to $\infty.$ Octahedral symmetry occurs
at the moment of closest approach.

Using (\ref{g4}) in the \Sk\ field ansatz, we find that the minimal
energy occurs at $c=1$, with the value $E=1.137.$ This is quite close
to the energy of the $B=4$ \sk\ $E=1.116$, and almost the same
as the energy of the best $B=4$ instanton generated Skyrme field
\cite{LM} which has $E=1.132.$

The Wronskian of the map (\ref{g4}) is proportional to $z(z^4-1)$ for
all values of $c$. This is the face polynomial of a cube, with faces
in the directions \hbox{$0,1,i,-1,-i,\infty $} (i.e. the directions of
the Cartesian axes). We understand from this why the baryon density
vanishes in these directions, and hence why the
\sk\ has a cubic shape, with its
energy concentrated on the vertices and edges of the cube (see Fig. 1c).\\

\subsection*{$N=5$}

The $B=5$ \sk\ of minimal energy has symmetry $D_{2d},$ which is
somewhat surprising. An octahedrally symmetric solution exists but has
higher energy \cite{BS3}. There is a family of rational maps
with two real parameters, with the generic map having $D_{2d}$
symmetry, but having higher symmetry at special parameter values.

The family of maps is
\be
R(z)=\frac{z(z^4+bz^2+a)}{az^4-bz^2+1}
\label{g5}
\ee
with $a$ and $b$ real. 
The two generators of the $D_{2d}$ symmetry are realized as
\be
R(i/\bar z)=i/\bar R(z) \hskip 1cm \mbox{and}  \hskip 1cm R(-z)=-R(z).
\ee
Additional symmetry occurs if $b=0$; $R(z)$ then has $D_4$ symmetry,
the symmetry
of a square. There is octahedral symmetry if, in addition, $a=-5$. This value ensures the $120^\circ$
rotational symmetry 
\be
R\bigg(\frac{iz+1}{-iz+1}\bigg)=\frac{iR(z)+1}{-iR(z)+1}.
\ee
The map $R(z)=z(z^4-5)/(-5z^4+1)$ has Wronskian
\be
W(z)=-5(z^8+14z^4+1)
\ee
which is proportional to the face polynomial of an octahedron.

Using the maps (\ref{g5}) in the \Sk\ field ansatz gives a structure
which is a polyhedron with eight faces. In the special case
$b=0,a=-5$, this polyhedron is an octahedron,
and the angular integral is
$\I=52.05$, whereas performing a numerical search over the parameters
$a$ and $b$ we find that $\I$ is minimized when $a=3.07$, $b=3.94$,
taking the value $\I=35.75.$ This is consistent with the
structure and symmetry of the known $B=5$ \sk, which is that of
a polyhedron made from four pentagons and four quadrilaterals.  
Minimizing over the profile function we find a Skyrme field
with energy $E=1.147$ (see Fig. 1d). 
There is a saddle point at the octahedral parameter
values, where $E=1.232$ (see Fig. 1j). There is a further, 
higher saddle point at
$a=b=0$, where the map (\ref{g5}) simplifies to $R(z) = z^5$,
and gives a toroidal Skyrme field.\\

\subsection*{$N=6$ and $N=8$}

The \sk s with $B=6$ and $B=8$ both have extended cyclic symmetry. It is
straightforward to find rational maps with these symmetries, and use them with our ansatz.

For $B=6$, the desired symmetry is $D_{4d}$. $D_4$ is generated by
$z\mapsto iz$ and $z\mapsto 1/z.$ The rational maps
\be
R(z)=\frac{z^4+ia}{z^2(iaz^4+1)}
\label{g6}
\ee
have this symmetry, since $R(iz)=-R(z)$ and $R(1/z)=1/R(z)$. If $a$ is
real $R(e^{i\pi/4}\bar{z})=i\bar{R}(z)$ and the rational maps have
$D_{4d}$ symmetry.  With these maps in our ansatz, the minimal energy
occurs at $a=0.16$, when $E=1.137$.  The Skyrme field has a polyhedral
shape consisting of a ring of eight pentagons capped by squares above
and below (see Fig. 1e).

For $B=8$, the symmetry is $D_{6d}$. $D_6$ is generated by $z\mapsto e^{i\pi/3}z$ and
$z\mapsto i/z.$ The rational maps
\be
R(z)=\frac{z^6-a}{z^2(az^6+1)}
\label{g8}
\ee have this symmetry. If $a$ is real they have $D_{6d}$ symmetry. This time the minimal energy \Sk\ 
field obtained using these maps has $E=1.118$ when $a=0.14.$ The
polyhedral shape is now a ring
of twelve pentagons capped by hexagons above and below (see Fig. 1g).\\

\subsection*{$N=7$}

In a sense, the $N=7$ case is similar to the cases $N=6$ and $N=8$,
but the \sk\ has dodecahedral shape. A dodecahedron is a
ring of ten pentagons capped by pentagons above and below.

Among the rational maps with $D_{5d}$ symmetry \be
R(z)=\frac{z^5-a}{z^2(az^5+1)} \ee the one with icosahedral symmetry
has $a=-3.$ The Wronskian is then proportional to the face polynomial
of a dodecahedron, namely $z(z^{10}+11z^5-1).$

In a different orientation, the icosahedrally symmetric map is
\be
R(z)=\frac{z^7-7z^5-7z^2-1}{z^7+7z^5-7z^2+1}
\ee
which has a similar structure to the octahedrally symmetric $N=4$
map ((\ref{g4}) with $c=1$). 

We have found it interesting to investigate the dodecahedron in yet another
orientation, where tetrahedral symmetry is manifest. There is a one parameter
family of degree seven maps with symmetry $T.$ The family is
\be
R(z)=\frac{bz^6-7z^4-bz^2-1}{z(z^6+bz^4+7z^2-b)}
\label{g7}
\ee 
where $b$ is complex. It is easy to verify that 
\be R(-z)=-R(z),
\hskip 0.3cm R(1/z)=1/R(z) \hskip 0.3cm \mbox{and} \hskip 0.3cm
R\bigg(\frac{iz+1}{-iz+1}\bigg)=\frac{iR(z)+1}{-iR(z)+1}.  
\ee 
For
imaginary $b$, the symmetry extends to $T_d$ and for real $b$ it
extends to $T_h$. When $b=0$ there is octahedral symmetry, and when $b=\pm
7/\sqrt{5}$ there is icosahedral symmetry $Y_h$. Using (\ref{g7}) in 
our ansatz, we have found the minimal energy at 
$b=\pm 7/\sqrt{5}$, which gives a dodecahedral \Sk\ field, with energy
$E=1.107$ (see Fig. 1f).
There is a saddle point at $b=0$ with a cubic shape.

The real $b$-axis represents an interesting dynamical process. If
we regard the rational maps as Jarvis maps of $N=7$ monopoles, then
motion along the real $b$-axis is a geodesic in which there is a
single monopole at the origin, and six monopoles approaching it along
the positive and negative Cartesian axes. They pass successively
through a dodecahedron, a cube and a second dodecahedron (rotated by
$90^\circ$ relative to the first) before separating into a
configuration similar to the incoming one (again rotated by
$90^\circ$, which affects the shape but not the positions of the
monopoles). A similar motion is possible with seven \sk s, but one
must allow for the varying potential energy. The energy of seven
separated \sk s is greater either than the cube or dodecahedron. If
the energy is sufficient, an oscillatory motion between the two
dodecahedra can occur, through the cubic configuration. At higher
energy there
can be a 7-\sk\ scattering process going through all these
configurations. 

\medskip
\medskip
So far we have used fairly {\sl ad hoc} methods to obtain our
symmetric rational maps.  However, this
approach becomes cumber\-some for higher degrees
and it is more efficient to use a 
systematic algorithm. By constructing symmetric rational maps we are
in effect computing bases for two-dimensional representations of
finite subgroups of $SU(2)$, for which classical group theory can be
employed. In the Appendix we describe the construction of symmetric
rational maps from this more systematic point of view. This
construction is illustrated with the example above, tetrahedral
symmetry for $N=7$, and some further examples occur below. \\

\subsection*{$N=9$}

Imposing tetrahedral symmetry on degree nine maps we find, see Appendix,
the one parameter family
\be
R(z)=\frac{5i\sqrt{3}z^6-9z^4+3i\sqrt{3}z^2+1+az^2(z^6-i\sqrt{3}z^4-z^2+i\sqrt{3})}{
z^3(-z^6-3i\sqrt{3}z^4+9z^2-5i\sqrt{3})+az(-i\sqrt{3}z^6+z^4+i\sqrt{3}z^2-1)}
\label{g9}
\ee where $a$ is real. This map is degenerate at the values
$a=1,-5,\pm\infty.$ In terms of the corresponding 9-monopole
configurations, the first of these degenerate values, $a=1$,
corresponds to a single monopole at the origin with eight monopoles on
the vertices of a cube at infinity. The value $a=-5$ corresponds to
four monopoles at infinity on the vertices of a tetrahedron, and
$a=\pm\infty$ represents six monopoles at infinity on the vertices of
an octahedron.

The angular integral $\I$ diverges as a degenerate map is
approached, so it is clear from the above that the family of maps
(\ref{g9}) contains at least three local minima for $\I$ as a function
of $a.$ In Fig. 4 we plot $\I$ as a function of $a$, from which it can
be seen that the global minimum occurs in the middle interval
$-5<a<1.$ More explicitly, the minimum occurs when $a=-1.98$, and the
energy of the resulting Skyrme field is $E=1.123.$ The Skyrme field
has a polyhedral shape consisting of four hexagons centred on the vertices of
a tetrahedron,
linked by four triples of pentagons (see Fig. 1h). \\

\subsection*{$N=17$}

For general $B>9$ the expected symmetries of the \sk\ are not great
enough to cut down the associated family of rational maps to just one
or two parameters.  Thus a minimization over a large family of
rational maps probably has to be undertaken.  However, given the
complicated nature of such a space of rational maps (for example,
recall the above one parameter family of degree nine maps, which
contains spurious local minima that lie in disconnected sectors) this
is a difficult numerical task.

Fortunately there are exceptional cases where we expect a highly
symmetric configuration to occur. One of these is at $B=17$,
where it has been conjectured \cite{BS2} that the Skyrmion has the
icosahedrally symmetric, buckyball structure of carbon 60.  An $N=17$
rational map with symmetry $Y_h$ is, see Appendix, \be
R(z)=\frac{17z^{15}-187z^{10}+119z^5-1}{z^2(z^{15}+119z^{10}+187z^5+17)}.
\label{m17}
\ee Using this map in our ansatz we obtain a \Sk\ field with energy
$E=1.092.$ The very low value for this energy 
supports the conjecture that $Y_h$ is the symmetry of the minimal energy
$B=17$ configuration.  The polyhedron does indeed have the buckyball
form (see Fig. 1i), consisting of twelve pentagons, each 
surrounded by five hexagons, making a total of 32 polygons.

From Fig. 2 it can be seen that the buckyball \sk\ is quite large,
and furthermore the profile function is extremely flat for small $r$.
This implies that there is a region inside the shell of the
buckyball where the \Sk\ field is close to the vacuum (in fact $U=-1$
but this is not important), possibly allowing 
smaller Skyrmions, for example $B=4$, to sit inside the buckyball with
little distortion. It would be interesting to investigate this
further. It may shed some light on the shell structure of
\sk s, which appears to be favoured over a crystal structure for the
cases investigated so far, but presumably fails for sufficiently large
$B$.\\

\subsection*{$N=11$}

We have already constructed several symmetric rational maps, such as
the $N=3$ example with axial symmetry and the $N=5$ example with octahedral
symmetry, which generate approximations to saddle point \Sk\ fields.
Although we have not computed the energy minimizing rational map of
degree eleven, we can compute an interesting saddle point map which has
icosahedral symmetry, see Appendix. Note that the existence of this
map proves the existence of an icosahedrally symmetric $N=11$
monopole, as conjectured in ref. \cite{HS2}. The rational map is \be
R(z)=\frac{11z^{10}+66z^5-1}{z(z^{10}+66z^5-11)}.  \ee The value of
the angular integral for this map is $\I=486.84$, which is very large;
it is even greater than the value for the $N=17$ map given above (see
Table 1).  This indicates that the minimal energy $B=11$ \sk\ will not
have $Y_h$ symmetry. Computing the energy we find $E=1.406$, which is
considerably higher than that of eleven well-separated $B=1$ \sk s.
This icosahedral configuration is shown in Fig. 1k.

\vbox{
\begin{center}
\begin{tabular}{|c|c|c|c|c|} \hline
B& $\I$ & APPROX & TRUE  & SYM\\
\hline
1 &  1.00 & 1.232 & 1.232  & $O(3)$ \\
2 &  5.81 & 1.208 & 1.171  & $O(2)\times\Z_2$ \\
3 &  13.58 & 1.184 & 1.143   & $T_{d}$ \\
4 & 20.65 & 1.137 & 1.116   & $O_h$ \\
5 & 35.75 & 1.147 & 1.116   & $D_{2d}$ \\
6 & 50.76 & 1.137& 1.109   & $D_{4d}$ \\
7 & 60.87 & 1.107 & 1.099   &  $Y_{h}$ \\
8 & 85.63 & 1.118 & 1.100   &  $D_{6d}$ \\
9 & 112.83 & 1.123 & 1.099   &  $T_{d}$ \\
17 & 367.41 &1.092 &1.073   & $Y_{h}$ \\
\hline
3* & 18.67 & 1.256 & 1.191 & $O(2)\times\Z_2$ \\
5* & 52.05 & 1.232 & 1.138 &  $O_{h}$ \\
11* & 486.84 & 1.406 & 1.158  & $Y_{h}$ \\
\hline
\end{tabular}
\end{center}
{\bf Table 1} : Comparison between the energies of approximate
Skyrmions generated from rational maps, and the energies of true
Skyrmions. The table gives the value of the angular integral $\I$, and
the associated Skyrme field energy (APPROX), together with the energy
of the true solution (TRUE), as determined in refs.  \cite{BS2,BS3},
and the symmetry (SYM) of the corresponding Skyrme field.  A $*$
denotes a saddle point configuration.}

\section{Rational Maps and Skyrmion Vibrations}
\news

Through our ansatz for \Sk\ fields in terms of rational maps we have
found approximations to several minimal energy \sk s of various baryon
numbers. It is natural to guess that varying the rational map
parameters will correspond to distortions of the \sk s into some of
their vibrational modes. It is interesting to investigate this, as
Barnes et al. \cite{BBT1,BBT2} have recently used a numerical
simulation of the \Sk\ equation to study the spectrum of vibrations
around the $B=2$ and $B=4$ Skyrmions. We can interpret some of the
qualitative features of their results in terms of rational maps, and
can predict what happens in some examples not yet analysed.

We consider first the vibrations of the $B=4$ \sk\ with octahedral
symmetry. This \sk\ has nine zero modes corresponding to translations,
rotations and isospin rotations. There are nine low-lying vibrational
modes, with frequencies somewhat less than the pion mass.  These modes
lie in multiplets transforming under certain irreducible
representations of the octahedral group $O$. In increasing order of
frequency, these representations are $E^O,A_2^O,F_2^O,F_2^O$ (in the notation
of ref.  \cite{Ha}), respectively of dimensions two, one, three and
three.  Barnes et al. have presented pictures of the \sk\ distortion
for these modes of vibration. The next mode is the breather mode (a
vibration of the scale size) which is invariant under the octahedral
group, and some higher frequency modes have been identified, separate
from the continuum of pion field vibrations.

It is the modes below the breather which can be identified with
variations of the rational map parameters. Recall that the rational
map of degree four with octahedral symmetry is 
\be
R_0(z)=\frac{z^4+2\sqrt{3}iz^2+1}{z^4-2\sqrt{3}iz^2+1}.
\label{R0}
\ee 
The general variation of this map, in which we preserve the
leading coefficient of the numerator as $1$ as a normalization, is 
\be
R(z)=\frac{z^4+\alpha z^3+(2\sqrt{3}i+\beta)z^2+\gamma z+1+\delta}{
  (1+\lambda)z^4+\mu z^3+(-2\sqrt{3}i+\nu)z^2+\sigma z+1+\tau} 
\ee
where $\alpha,\beta,\gamma,\delta,\lambda,\mu,\nu,\sigma,\tau$ are
small complex numbers. We now calculate the effect of the
transformations of the octahedral group leaving $R_0$ fixed. For
example, the $90^\circ$ rotation, represented by the transformation
$R(z)\mapsto 1/R(iz)$ leaves $R_0$ fixed, but transforms the
general map $R(z)$ to 
\be 
\widetilde R(z)=\frac{(1+\lambda)z^4-i\mu
  z^3+(2\sqrt{3}i-\nu)z^2+i\sigma z+1+\tau} {z^4-i\alpha
  z^3-(2\sqrt{3}i+\beta)z^2+i\gamma z+1+\delta}.  
\ee 
Normalizing this
by dividing top and bottom by $1+\lambda$, and ignoring quadratic and
smaller terms in the small parameters, we get 
\be 
\widetilde
R(z)=\frac{z^4-i\mu z^3+(2\sqrt{3}i-\nu-2\sqrt{3}i\lambda)z^2+i\sigma
  z+1+\tau-\lambda}{(1-\lambda)z^4-i\alpha
  z^3+(-2\sqrt{3}i-\beta+2\sqrt{3}i\lambda)z^2+i\gamma
  z+1+\delta-\lambda}.  
\ee 
Therefore the transformation acts linearly on the
nine parameters $\alpha,..,\tau$ via a $9\times 9$ matrix
that can be read off from this expression. The
only contribution to the trace of the $9\times 9$ matrix is the 
$-1$ associated with the replacement
of $\lambda$ by $-\lambda$ in the leading term of the denominator. So
the character $\chi$ of the $90^\circ$ rotation in this
representation is $-1$.

We really need to consider this representation as a real
eighteen-dimensional one, so the character above becomes $\chi = -2$.
From now on we shall work with real representations.

Similar calculations for the elements of each conjugacy class of the
octahedral group give the characters listed in Table 2, where $I$ is the
identity, $C_4$ denotes a $90^\circ$ rotation and $C_4^2$ is the
square of this, $C_3$ denotes a $120^\circ$ rotation and $C_2$ a
rotation by $180^\circ$ which is not the
square of a $90^\circ$ rotation.\\

\vbox{
\begin{center}
\begin{tabular}{|c|c|} \hline
Class & Character $\chi$\\
\hline
$I$& 18\\
$6C_4$ & $-2$\\
$3C_4^2$ & 2 \\
$8C_3$ & 0\\
$6C_2$ &$-2$ \\
\hline
\end{tabular}
\end{center}
{\bf Table 2} : Characters of the group $O$ acting on the real 
eighteen-dimensional parameter space of
deformations of the octahedral degree four rational map.}\ 
The character table of $O$ tells us that this eighteen-dimensional representation
 splits into the irreducible components 
$ 2A_2^O+ 2E^O+ 2F_1^O+ 2F_2^O.$

To find which of the irreducible representations correspond to true
vibrations we need to remove those corresponding to zero modes. First
we need to remove the representation associated with $SU(2)$ M\"obius
transformations of $R_0(z)$ which correspond to isospin rotations of the
\Sk\ field.  So we consider the infinitesimal deformations 
\be
R_0(z)\mapsto
\frac{(1+i\epsilon)R_0(z)+\epsilon'}{-\bar\epsilon'R_0(z)+(1-i\epsilon)} 
\ee
where $\epsilon$ is real, and $\epsilon'$ complex. Under the
transformations of the octahedral
group the characters are $\chi(I)=3, \ \chi(C_4)=-1, \ 
\chi(C_4^2)=3, \ \chi(C_3)=0, \ \chi(C_2)=-1$, so these parameter
variations transform as $A_2^O+
E^O$.  Similarly, the parameter variations
which correspond to translations and rotations transform under the
octahedral group as $F_1^O+ F_1^O$. From the above eighteen-dimensional
representation we therefore subtract $ A_2^O+ E^O+ F_1^O+ F_1^O$ to obtain the
representation of the true vibrations, which has the irreducible components
$A_2^O+ E^O+ F_2^O+ F_2^O$, and is nine-dimensional. These irreducible
representations are precisely the ones found by Barnes et al. for the
low-lying \sk\ vibrations.

Barnes et al., in their calculations of the vibrations of the
$B=2$ toroidal \sk \cite{BBT2},
found just one doubly degenerate mode of vibration of low frequency (below the breather),
and it corresponds to the deformation of the rational map (\ref{g2}) as $a$ varies away from
zero (corresponding to the separation mode for two monopoles).

We have done a similar analysis for the vibrational modes of the $B=3$
tetrahedral \sk. From the rational map parametrization we predict that
there are
five low-lying modes, transforming as $E^T+ F^T$ of the tetrahedral group
$T_d.$ This result slightly disagrees with Walet's \cite{Wa} estimate
of the vibrations using the instanton approximation of \Sk\ fields.
Although Walet found the lowest modes to be in an $E^T+F^T$, he also
found a second triplet of modes just below the breather.  Our results
suggest that this second triplet should really have a higher
frequency, but this must be checked using the exact solution and its
vibrations.

Since the $B=7$ Skyrmion has $Y$ symmetry, its vibrational modes also fall
into large degenerate multiplets. The rational maps involved
have degree seven and it is useful to simplify the calculation by
adopting the representation theory perspective of the Appendix. 

A degree $N$ rational map $R=(p_0,q_0)$ is $G$-symmetric when $p_0$
and $q_0$ span a two-dimensional representation of $G$ inside
$\underline{N+1}$.  This means that acting with $g\in G$ on $(x,y)$
has the effect of transforming $(p_0,q_0)$ by some $2\times 2$ matrix
$D_g$. Put another way, the $g$ transformation of $(x,y)$ followed by the
$D^{-1}_g$ transformation of the rational map leaves $(p_0,q_0)$
unchanged.  To find the transformation properties of the vibrations, a
general $(p,q)$ is transformed in this way.

We know how a general homogeneous polynomial transforms under $G$; it
is in the representation $\underline{N+1}|_G$. We also know the $D_g$
representation; it is the two-dimensional representation in
$\underline{N+1}|_G$ corresponding to $R$. The $D_g^{-1}$
representation $E$ can be calculated from this. Transforming $p$ and
$q$ under $\underline{N+1}|_G$ and then under $E$ is a
$\underline{N+1}|_G\times E$ transformation of $(p,q)$, where $(p,q)$
is regarded as a $(2N+2)$-dimensional vector. Thus, to find the
transformation properties of the vibrations we decompose
$\underline{N+1}|_G\times E$ into irreducible representations of $G$.

In the $B=7$ case 
\be 
\underline{8}|_Y=E^{\prime Y}_2+I^{\prime Y}.
\ee 
The icosahedral Skyrmion corresponds to $E^{\prime Y}_2$. That is
the representation of the $D_g$'s mentioned above. All elements of $Y$
lie in the same conjugacy class as their inverses, so the
$D_g^{-1}$ representation is also $E^{\prime Y}_2$. Each character of
$\underline{8}|_Y\times E^{\prime Y}_2$ is obtained by multiplying the
corresponding one for $\underline{8}|_Y$ with that for $E^{\prime
  Y}_2$.  These are listed in Table 3.

\vbox{
\begin{center}
\begin{tabular}{|c|c|c|c|} \hline
Class & $\underline{8}|_Y$ 
      & $E^{\prime Y}_2$
      & $\underline{8}|_Y\times E^{\prime Y}_2$ \\
\hline
$I$        & 8               &  2               &16                \\
$12C_5$    &$-1/2-\sqrt{5}/2$&$ 1/2-\sqrt{5}/2$ &1   \\
$12C_5^2$  &$ 1/2-\sqrt{5}/2$&$-1/2-\sqrt{5}/2$ &1\\
$20C_3$    & 1               &1                 &1                 \\
$15C_2$    & 0               &0                 &0                 \\
\hline
\end{tabular}
\end{center}
{\bf Table 3} : Characters for representations of $Y$ associated with
vibrations of the $B=7$ Skyrmion.}\ \\

Knowing the characters, we find the decomposition \be
\underline{8}|_Y\times E^{\prime Y}_2=A^Y+F_1^Y+F_2^Y+G^Y+H^Y.\ee
There are copies of this decomposition corresponding to real
variations and to imaginary variations. This means the variations
around the $B=7$ Skyrmion transform as $2A^Y+2F_1^Y+2F_2^Y+2G^Y+2H^Y$.
The $2A^Y$ are the trivial variations caused by multiplying the
icosahedral $p_0$ and $q_0$ by the same constant.
The vector representation of the icosahedral group is
$F_1^Y$, so translations and rotations account for $2F_1^Y$, and 
M\"obius transformations account for an $F_2^Y$. The representation of
the true vibrations therefore has
irreducible components $F_2^Y+2G^Y+2H^Y$, with degeneracies three,
four, four, five and five, respectively.

\section{Morse Function on Monopole Moduli Spaces}
\news

The \Sk\ field ansatz (\ref{ansatz}), using a rational map $R(z)$,
leads to a contribution to the \Sk\ energy given by
\be
\I=\frac{1}{4\pi}\int
 \bigg(
\frac{1+\vert z\vert^2}{1+\vert R\vert^2}
\bigg\vert\frac{dR}{dz}\bigg\vert\bigg)^4 \frac{2i \ dz  d\bar z }{(1+\vert z\vert^2)^2}.
\label{ii}
\ee

Now we may regard $\I$ simply as a function on the space of rational
maps of any given degree, $N$.  If we also identify rational maps with
monopoles, via the Jarvis construction, $\I$ becomes a function on the
$N$-monopole moduli space. $\I$ respects some, but not all,
the natural symmetries of the monopole moduli space. $\I$ is invariant
under rotations of the target $S^2$, hence descends to the usual
$(4N-1)$-dimensional moduli space ${\cal M}_N$. It is also
invariant under rotations of the domain $S^2$, hence is unchanged when
the monopole configuration is rotated. $\I$ is not, however, invariant
under a translation of the monopole configuration in $\R^3.$

It appears that $\I$ is a \lq\lq proper\rq\rq\ Morse function, that
is, the set of rational maps, and hence monopoles, for which $\I$ has
any particular finite value is compact.  We have not
verified this in general. It is necessary to prove that $\I$ tends to
infinity whenever the rational map degenerates. We have seen this
happen in several cases mentioned in Section 4. Such a degeneracy
corresponds to one or more monopoles moving off to infinity.

We have calculated one special case analytically. Consider the rational maps
$R(z)=cz.$ The phase of $c$ is unimportant, so let $c$ be real and
positive. $R$ degenerates if either $c$ becomes zero or infinite.
Since $R$ has degree one, it is the Jarvis map of a single monopole,
centred, in fact, at
$(0,0,2\log c).$ The integral $\I$ reduces to \be \I=2c^4\int_0^\infty
\frac{\rho(1+\rho^2)^2 \ d\rho}{(1+c^2\rho^2)^4}
=\frac{1}{3}(c^2+1+1/c^2).  \ee So $\I$ indeed diverges if
$c\rightarrow 0$ or $c\rightarrow\infty.$ The minimal value is $\I=1$
when $c=1$, as expected.  For a rational map of the form
$R=(z-a)/(z-b)$, the integral again diverges as $b$ approaches $a$;
this is equivalent, by a M\"obius transformation, to the example
$R(z)=cz$ with $c\rightarrow\infty.$ Generally, one may expect $\I$ to
diverge whenever a zero and a pole of $R$ come together.

Having a proper Morse function $\I$ defined on the monopole moduli
space helps us understand the topology of the moduli space. We have
investigated the 3-monopole moduli space in this way. The
stationary points of $\I$ on ${\cal M}_3$ consist of a number of
orbits of the rotation group $SO(3)$. Among
the $D_2$ symmetric maps of the form (\ref{g3})
we found just two types of stationary point for $\I$.
Assuming that $\I$ has no further types of stationary point, then on
${\cal M}_3$, $\I$ has two stationary orbits. One
is the set of $N=3$ tori (centred at the origin). This is a
two-dimensional orbit. Each torus is a saddle
point, with two independent unstable modes (related by rotations about
the symmetry axis).  The unstable manifold of this orbit (suitably
completed) is therefore
a 4-cycle. By symmetry, the unstable manifold includes all the
rational maps (\ref{g3}) with $a$ lying on the great circle segment
$0\le ia <1.$ The unstable manifold therefore consists of the orbits
under $SO(3)$ of all the rational maps of the form (\ref{g3}), with
$a$ in this interval. The other stationary orbit is the set of
tetrahedra (again centred at the origin), which is the orbit of
minima. This orbit is
three-dimensional, and completes the 4-cycle.  We have tried to
visualize this 4-cycle as a smooth submanifold of ${\cal M}_3$, but
have found this difficult in the neighbourhood of the tetrahedra.

A 4-cycle is the basic non-trivial compact homology cycle which is
predicted by the calculations of Segal \& Selby \cite{SS}. It would be
interesting if the Sen 4-form, representing a bound state of three
monopoles \cite{Se}, were concentrated around the particular 4-cycle
we have found.

These calculations suggest that further investigation of
$\I$ as a Morse function on ${\cal M}_N$ would be worthwhile.

\section{Conclusion}
\news 
We have introduced a new ansatz for \Sk\ fields, based on
rational maps. This allows us to construct good approximations 
to several \sk s and helps us understand
the similarities which have been observed between \sk s and BPS
monopoles. A certain black hole with hair
has states with a remarkably similar structure to Skyrmions, also related to
rational maps \cite{RW}. Thus it appears that
a whole class of solitonic objects in three space
dimensions may be understood via the kind of rational map approach
which we employ here.

We have used our ansatz to study the low-lying vibrational modes 
of \sk s. For the $B=2$ and $B=4$ Skyrmions,
our results agree qualitatively with those obtained numerically, and
we can predict the structure of the vibrational spectrum for other
cases, in particular $B=3$ and $B=7$.

Finally, the relationship between 
monopoles and \sk s has led us to an interesting Morse
function on the monopole moduli spaces which may be of use in
understanding the homology of the moduli spaces and thus
predictions made by duality.

\section*{Acknowledgements}

It is a pleasure to thank Erick Weinberg for fruitful discussions at
an early stage of this work. We also thank Richard Battye and Neil
Turok for useful comments.  NSM and PMS are grateful for the
hospitality of the Isaac Newton Institute, where some of this work was done. 
PMS acknowledges support from the Nuffield Foundation and thanks the Cambridge
Philosophical Society for a travel bursary.  CJH thanks the EPSRC for
a research studentship and the British Council for a Chevening award.
\appendix
\section*{Appendix \ \ Systematic calculation of symmetric maps}
\news
\renewcommand{\theequation}{A\arabic{equation}}
\ \indent

For low degrees, symmetric rational maps may be constructed by
explicitly performing the group transformations on a general rational
map and deriving constraints on the coefficients. For higher degrees
and for larger groups it is useful to employ the theory of group
representations in the construction of the symmetric rational maps. In
this Appendix such a construction will be described, and applied to
the example of degree seven maps with tetrahedral symmetry.

To construct symmetric rational maps it is convenient to employ
homogeneous projective coordinates $x$ and $y$ on the Riemann sphere,
rather than the inhomogeneous $z=x/y$ employed earlier.  A rational
map is a map from Riemann sphere to Riemann sphere of the form 
\be
R(x,y)=(p(x,y),q(x,y))
\ee 
where $p$ and $q$ are homogeneous polynomials. 
In the $(x,y)$ coordinates, an $SO(3)$ rotation
in space by
$\theta$ about the direction of the unit vector $(n_1,n_2,n_3)$ is
realized by the
$SU(2)$ transformation $\exp\left(i\frac{\theta}{2}
{\bf n}\cdot{\mbox{\boldmath $\sigma$}} \right)$, whose action on the
Riemann sphere is
\bea
x\mapsto x^{\prime}&=&(d+ic)x-(b-ia)y\label{Mtrans}\\
y\mapsto y^{\prime}&=&(b+ia)x+(d-ic)y\nonumber 
\eea 
where $a=n_1\sin{\frac{\theta}{2}}$, $b=n_2\sin{\frac{\theta}{2}}$,
$c=n_3\sin{\frac{\theta}{2}}$ and $d=\cos{\frac{\theta}{2}}$.
Furthermore, for our purposes, two rational maps are equivalent if
they can be mapped into each other by an $SU(2)$ transformation of
the target sphere, that is by a transformation of $p$ and $q$ of the
form (\ref{Mtrans}). A rational map is symmetric under some finite
group $G \subset SU(2)$ if $G$ transformations of $x$ and $y$ map it
into an equivalent map.

A degree $N$ homogeneous polynomial is a polynomial of the form 
\be
p(x,y)=\sum_{i=0}^N a_ix^iy^{N-i}.
\ee 
Under $SU(2)$ transformations
(\ref{Mtrans}) of $x$ and $y$ the space of degree $N$ homogeneous
polynomials transforms under the unique irreducible $(N+1)$-dimensional
representation of $SU(2)$: $\underline{N+1}$.  This $\underline{N+1}$
is also a representation of any finite subgroup $G$ of $SU(2)$,
generally reducible. It is easy to
calculate its decomposition into irreducible representations,
because, in $\underline{N+1}$, 
the element $\exp\left(i\frac{\theta}{2}
{\bf n}\cdot{\mbox{\boldmath $\sigma$}} \right)$ has character
\be\frac{\sin\left(\frac{N+1}{2}\right)\theta}{\sin\frac{\theta}{2}}\ee
for any ${\bf n}$.
There are tables of these reductions given in, for example, ref.
\cite{Ketal}.

Suppose two degree $N$ homogeneous polynomials $p(x,y)$ and $q(x,y)$
lie in the same two-dimensional representation of $G$; then, $G$
transformations of $x$ and $y$ will result in $GL(2,\C)$
transformations of $(p(x,y),q(x,y))$. If, further, $p(x,y)$ and
$q(x,y)$ are orthonormal as vectors in the $\underline{N+1}$ carrier
space, then, projectively, the $G$ action on $x$ and $y$ results only
in $SU(2)$ transformations of $(p(x,y),q(x,y))$.  Therefore, the
rational map $R(x,y)=(p(x,y),q(x,y))$ is $G$ symmetric.

This means that there is a systematic way of deciding whether there
are $G$ symmetric maps of some degree $N$.
The representation $\underline{N+1}$ is decomposed into irreducible
representations of $G$. If 
\be 
\underline{N+1}|_G=E+\mbox{other irreducible representations of}\;G,
\label{gendecomp}
\ee
where $E$ is a two-dimensional irreducible
representation of $G$, and if 
the basis polynomials for $E$
have no common factor, then there is a $G$ symmetric degree $N$ map. If
they have a common factor then the resulting rational map has lower degree.
This occurs when the $E$ in
$\underline{N+1}$ is the product of lower degree
polynomials; this is illustrated with an example below.  It might
also happen that 
\be \underline{N+1}|_G=A_1+A_2+\mbox{other
  irreducible representations of}\;G,
\label{gendecomp2}
\ee
where $A_1$ and $A_2$ are one-dimensional representations of $G$. In
this case there is a one parameter family of $G$ symmetric rational
maps: if $p(x,y)$ is in $A_1$ and $q(x,y)$ is in $A_2$ then the family \be
R(x,y)=(ap(x,y),q(x,y))\ee is $G$ symmetric.

The example of tetrahedral symmetry for degree seven is now discussed.
Let us consider the representation $\underline{8}$. Under restriction to $T$
\be 
\underline{8}|_{T}=2E^{\prime T}+G^{\prime T},\label{8toT}
\ee
that is, two two-dimensional irreducible representations of $T$ occur
in the decomposition of $\underline{8}$.
Furthermore, there is an arbitrariness in the decomposition 
\be 
2E^{\prime T}=E^{\prime T}+E^{\prime T},
\ee
and this allows a one parameter family of tetrahedrally symmetric
rational maps to be constructed.

The tetrahedral group is both a subgroup of the octahedral
group $O$ and a subgroup of the icosahedral group $Y$.
We can decompose $\underline{8}$ as a representation of $Y$ and
of $O$. We find
\begin{eqnarray}
  \underline{8}|_O&=&E^{\prime O}_{1}+E^{\prime
  O}_2+G^{\prime O}\label{8toO},\\
  \underline{8}|_Y&=& E_2^{\prime Y}+
  I^{\prime Y}. \label{8toY}\end{eqnarray}    
We can decompose these representations further by restriction to
$T$
\begin{eqnarray}E_1^{\prime O}|_T&=&E^{\prime T},\label{Odecom}\\
                E_2^{\prime O}|_T&=&E^{\prime T},\nonumber\\
                  G^{\prime O}|_T&=&G^{\prime T} \nonumber\end{eqnarray}
and
\begin{eqnarray}
                E_2^{\prime Y}|_T&=&E^{\prime T},\\
                  I^{\prime Y}|_T&=&E^{\prime T}+G^{\prime T}.
\nonumber\end{eqnarray}
In this way, we see that $T$ has two identical two-dimensional irreducible
representations in $\underline{8}$. $O$ has two as well but they are
different and $Y$ only has one.  The carrier spaces of these
representations are two-dimensional subspaces of the carrier
space of $\underline{8}$, a space which is realised as degree seven
homogeneous polynomials. The symmetric rational maps we wish to
calculate are constructed from the bases of the two-dimensional
spaces.

There are simple and venerable methods for calculating such bases
explicitly. They are explained in Serre's book \cite{S}.
Consider $U$, a reducible representation of a group $G$,
\begin{eqnarray} 
G&\rightarrow& GL(U)\label{repU}\\g&\mapsto&\rho(g),\nonumber\end{eqnarray}
which decomposes into irreducible representations $V_i$,
\begin{eqnarray}
  U&=&V_1+\ldots+ V_1+ V_2+ \ldots +
  V_2+ \ldots\ldots+
  V_h+\ldots+ V_h \label{decompU}\\
   &=&W_1+\ldots+ W_h\nonumber
\end{eqnarray}
where
\be 
W_i=V_i + V_i + \ldots + V_i. 
\ee 
If the irreducible representation $V_i$ has character $\chi_i(g)$ for
$g\in G$, and $n_i=\mbox{dim}W_i$, then
\be 
P_i=\frac{n_i}{|G|}\sum_{g\in G}\chi_i(g)^{\star}\rho(g)
\label{projfor}
\ee
is the projection operator
\be P_i:U\rightarrow W_i.\ee
Using {\scriptsize MAPLE} these projection operators can be calculated.

Since $E^{\prime T}$ appears twice in ${\underline 8}|_T$, projection
onto $E^{\prime T}$ gives a four-dimensional space. To work out a
basis for this space, the projection operator 
\be
P:\underline{8}\rightarrow 2E^{\prime T} \ee 
must be calculated using (\ref{projfor}). The $T\subset SU(2)$
transformations of $(x,y)$ are first calculated explicitly.
In the orientation where
each edge of the tetrahedron has its midpoint on a Cartesian axis, the
$C_2$ element about the $x_3$-axis has $c=-1$ and $a=b=d=0$ and hence
\bea
&&x^{\prime}=-ix\\
&&y^{\prime}=iy.\nonumber
\eea
The $C_3$ element about the $x_1=x_2=x_3$ axis has $a=b=c=d=1/2$
and hence
\bea
&&x^{\prime}=\frac{1+i}{2}x+\frac{1-i}{2}y\\
&&y^{\prime}=-\frac{1+i}{2}x+\frac{1-i}{2}y.\nonumber
\eea
These two generate $T$, so we can calculate expressions for
the $(x,y)$ transformations for all 24 elements of $T$. 
Using {\scriptsize MAPLE}, we calculate the effects of these
transformations on degree
seven polynomials, hence determining the $8 \times 8$ matrices 
$\rho(g)$ for each element $g \in T$, and hence, using (\ref{projfor}), 
the projection operator $P$. 
The resulting polynomials in the image of $P$ are
\bea
&&p_1(x,y)=-7x^4y^3-y^7,\label{W}\\
&&p_2(x,y)=x^7+7x^3y^4,\nonumber\\
&&p_3(x,y)=x^6y-x^2y^5,\nonumber\\
&&p_4(x,y)=x^5y^2-xy^6.\nonumber \eea 
This particular basis is chosen because it is convenient for what follows.

From (\ref{Odecom}) it follows that there lie in this
four-dimensional space two different representations of the octahedral
group $O$. In the chosen orientation, $O$ is generated by $T$ and the
$C_4$ rotation around the $x_3$-axis: \bea
&&x^{\prime}=\frac{1+i}{\sqrt{2}}x\\
&&y^{\prime}=\frac{1-i}{\sqrt{2}}y\nonumber \eea and so the projection
operators for $E_1^{\prime O}$ and $E_2^{\prime O}$ can be calculated.
It is found that $p_1(x,y)$ and $p_2(x,y)$ are a basis for
$E_1^{\prime O}$ and $p_3(x,y)$ and $p_4(x,y)$ are a basis for
$E_2^{\prime O}$. The rational map \be R(x,y)=(p_1(x,y),p_2(x,y)) \ee
is therefore octahedrally symmetric. However $p_3(x,y)$ and $p_4(x,y)$
have a common factor and the corresponding rational map is spurious;
it is not of degree seven.  This is not surprising. The
one-dimensional representation $A_2^O$ in
$\underline{7}|_{O}=A_2^O+F_1^O+F_2^O $ has basis $x^5y-xy^5$, the
two-dimensional representation ${\underline 2}|_{O}=E_1^{\prime O}$
has basis $x$, $y$, and $A_2^O\times E_1^{\prime O}=E_2^{\prime O}$.

Recall that $T$ is also a subgroup of $Y$. In fact, for our choice of
orientation for the tetrahedral group, there are two possible
icosahedral groups with it as a subgroup. The group $Y$ is generated
by $T$ and a $C_5$ element. The two choices of $Y$ correspond to
adding a $C_5$ rotation about the radial line passing through
$(-1,0,\tau)$ or about the line passing through $(1,0,\tau)$, where
$\tau=(1+\sqrt{5})/2$. 
The two possibilities are related by a rotation
by $90^{\circ}$ about the $x_3$-axis.  The $E_2^{\prime Y}$ has basis
$p_1(x,y)\pm (7/\sqrt{5})p_3(x,y)$ and $p_2(x,y)\pm (7/\sqrt{5})p_4(x,y)$;
the sign depends on the choice of $C_5$ element.

Let us now consider the decomposition of $2E^{\prime T}$ into
$E^{\prime T}+E^{\prime T}$. Luckily, such decompositions are
discussed in \cite{S} where the following construction is presented.
We have, generally, some reducible representation $U$,
where, as in (\ref{decompU}),
\be 
U=W+\mbox{other irreducible representations of }G 
\ee
and $W$ is the sum of $m$
identical irreducible representations $V$, 
\be
W=mV.\label{decomp}\\
\ee 

Let $n=\mbox{dim} V$ (in our example $n=2$). In
$V$ each $g\in G$ is represented by an $n\times n$ matrix, say $r(g)$.
From these the projection operators
\be
P_{\alpha\beta}=\frac{n}{|G|}\sum_{g\in G}
r_{\alpha\beta}(g^{-1})\rho(g)
\ee 
are calculated. Here, $\alpha$, $\beta$ are simply the matrix indices
of $r$. Now $P_{\alpha\alpha}$ projects onto an $m$-dimensional
space we will call $\Omega_{\alpha}$, and $W$ can be expressed as the
direct sum
\be 
W=\Omega_1+\Omega_2+\ldots+ \Omega_n .
\ee 
Furthermore, the map $P_{\beta\alpha}$ is an isomorphism from
$\Omega_{\alpha}$ to $\Omega_{\beta}$ and vanishes on all $\Omega_{\gamma}$ for
$\gamma\not=\alpha$. If $(\omega_1,\omega_2,\ldots,\omega_m)$ is a
basis for $\Omega_1$ then the space spanned by
\be
Y_{\nu} =
(\omega_{\nu},P_{21}(\omega_\nu),P_{31}(\omega_{\nu}),\ldots,
P_{n1}(\omega_\nu)).
\ee 
is isomorphic to $V$ and 
\be 
W=Y_1+Y_2+\ldots+ Y_m
\ee 
is a decomposition of $W$ of the form (\ref{decomp}). Choosing a
particular decomposition is equivalent to choosing a particular basis
$(\omega_1,\omega_2,\ldots,\omega_m)$ for the space $\Omega_1$.

In the example we are considering, $W=2E^{\prime T}$. This
space is spanned by the polynomials (\ref{W}). Using  {\scriptsize
  MAPLE} the projection operators $P_{11}$ and $P_{21}$ are
constructed. It is found that the space $P_{11}:W\rightarrow \Omega_1$ is
spanned by $p_1$ and $p_3$. Choosing a vector $p_1+bp_3$ in this space
defines a particular  $E^{\prime T}\subset 2E^{\prime T}$. Using
$P_{21}$ we derive from this the one-parameter family of tetrahedrally
symmetric rational maps
\be R(x,y)=(p_1+b p_3,p_2+b p_4),\ee
or in inhomogenous coordinates
\be
R(z)=\frac{bz^6-7z^4-bz^2-1}{z(z^6+bz^4+7z^2-b)}
\ee 
where $b$ is complex. For imaginary $b$, the symmetry extends to $T_d$
and for real $b$ it extends to $T_h.$ For $b=0$, there is
octahedral symmetry $O_h$ and for $b=\pm 7/\sqrt{5}$ there is icosahedral symmetry
$Y_h.$ 

We have used similar methods to calculate icosahedrally symmetric maps
for degrees eleven and seventeen and to calculate tetrahedrally
symmetric maps of degree nine. In the two icosahedral cases there is a
single symmetric rational map 
\bea
&&\underline{12}|_{Y}= E_1^{\prime Y}+G^{\prime Y}+I^{\prime Y},\\
&&\underline{18}|_{Y}= E_2^{\prime Y}+G^{\prime Y}+2I^{\prime
  Y},
\nonumber
\eea 
and to construct the map we need only
calculate a basis for $E^{\prime Y}$ in each case. For degree nine
\be \underline{10}|_{T}= E^{\prime T}+2G^{\prime T}.  \ee The
representation $G^{\prime T}$ is a sum of two two-dimensional
irreducible representations of $T$. Because they are complex conjugate
representations they are amalgamated under the name $G^{\prime T}$ in
the standard nomenclature. If we write $G^{\prime T}=E_1^{\prime
  T}+E_2^{\prime T}$ then \be \underline{10}|_{T}= E^{\prime
  T}+2E_1^{\prime T}+2E_2^{\prime T} \ee and a one parameter family of
symmetric rational maps can be constructed from $2E_1^{\prime T}$. The
corresponding family constructed
from $2E_2^{\prime T}$ is related by inversion. The representation 
$E^{\prime T}$ does not give a genuine degree nine map.

\newpage
\noindent{\bf Figure Captions}

\ \\

\noindent{\bf Fig. 1:} 
Surfaces of constant baryon density for the following \Sk\ fields:

a) $B=2$ torus 

b) $B=3$ tetrahedron

c) $B=4$ cube

d) $B=5$ with $D_{2d}$ symmetry

e) $B=6$ with $D_{4d}$ symmetry

f) $B=7$ dodecahedron

g) $B=8$  with $D_{6d}$ symmetry

h) $B=9$ with tetrahedral symmetry

i) $B=17$ buckyball

j) $B=5$ octahedron

k) $B=11$ icosahedron\\

\noindent{\bf Fig. 2:} The profile functions $f(r)$ for baryon numbers 
one to nine and also seventeen.\\

\noindent{\bf Fig. 3:} The $a$-sphere parametrizing the degree three rational
maps (\ref{g3}). Crosses denote degenerate maps, dots denote toroidal
maps and triangles denote the tetrahedral maps.\\

\noindent{\bf Fig. 4:} The integral $\I$ for the family of degree nine
maps (\ref{g9}).\\

\begin{figure}[p]
\begin{center}
\vskip -1cm
\epsfig{file=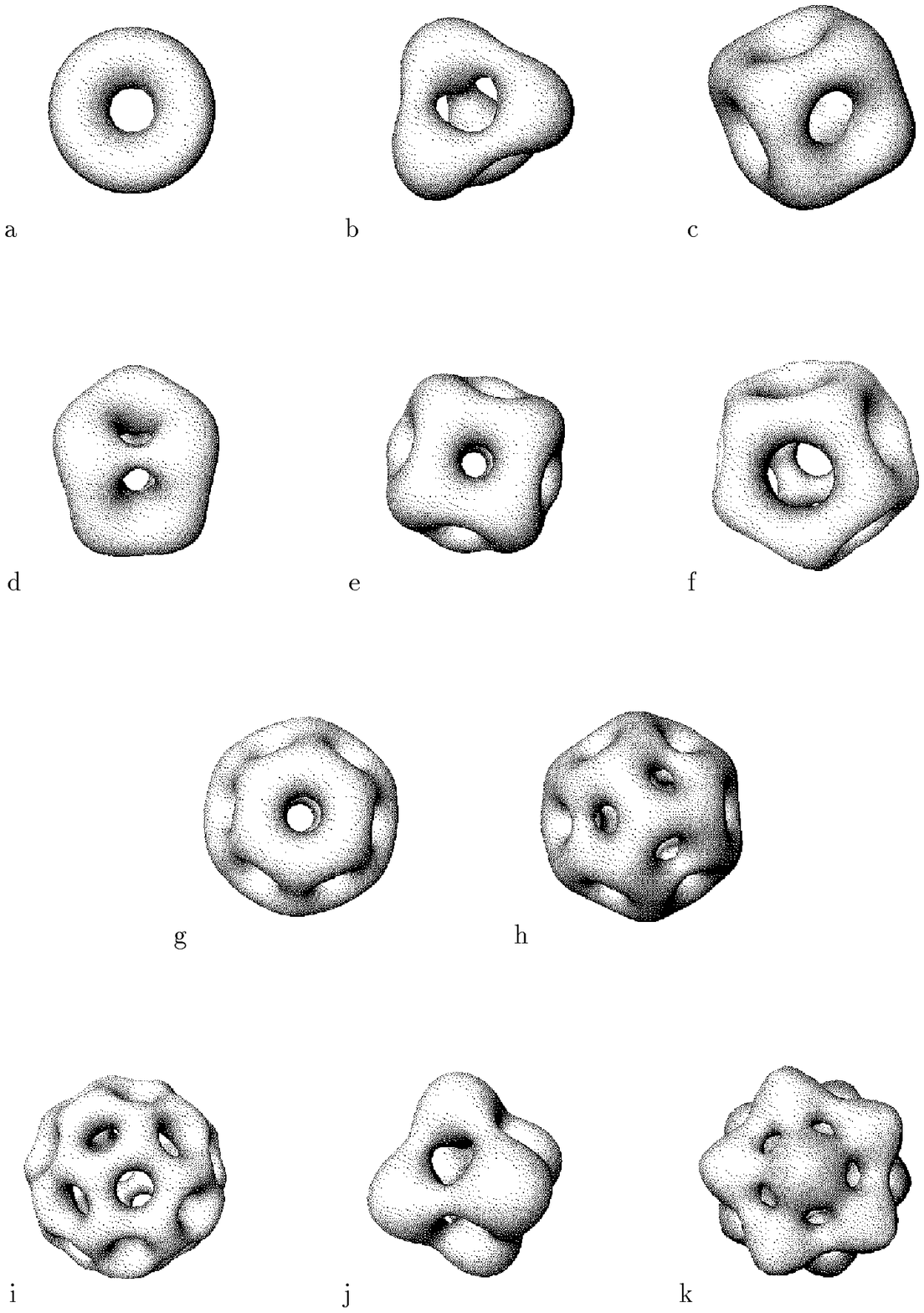,bbllx=70pt,bblly=160pt,bburx=490pt,bbury=750pt,
width=15cm}\\[10pt] 
{\bf Figure 1}
\end{center}
\end{figure}

\begin{figure}[p]
\begin{center}
\epsfig{file=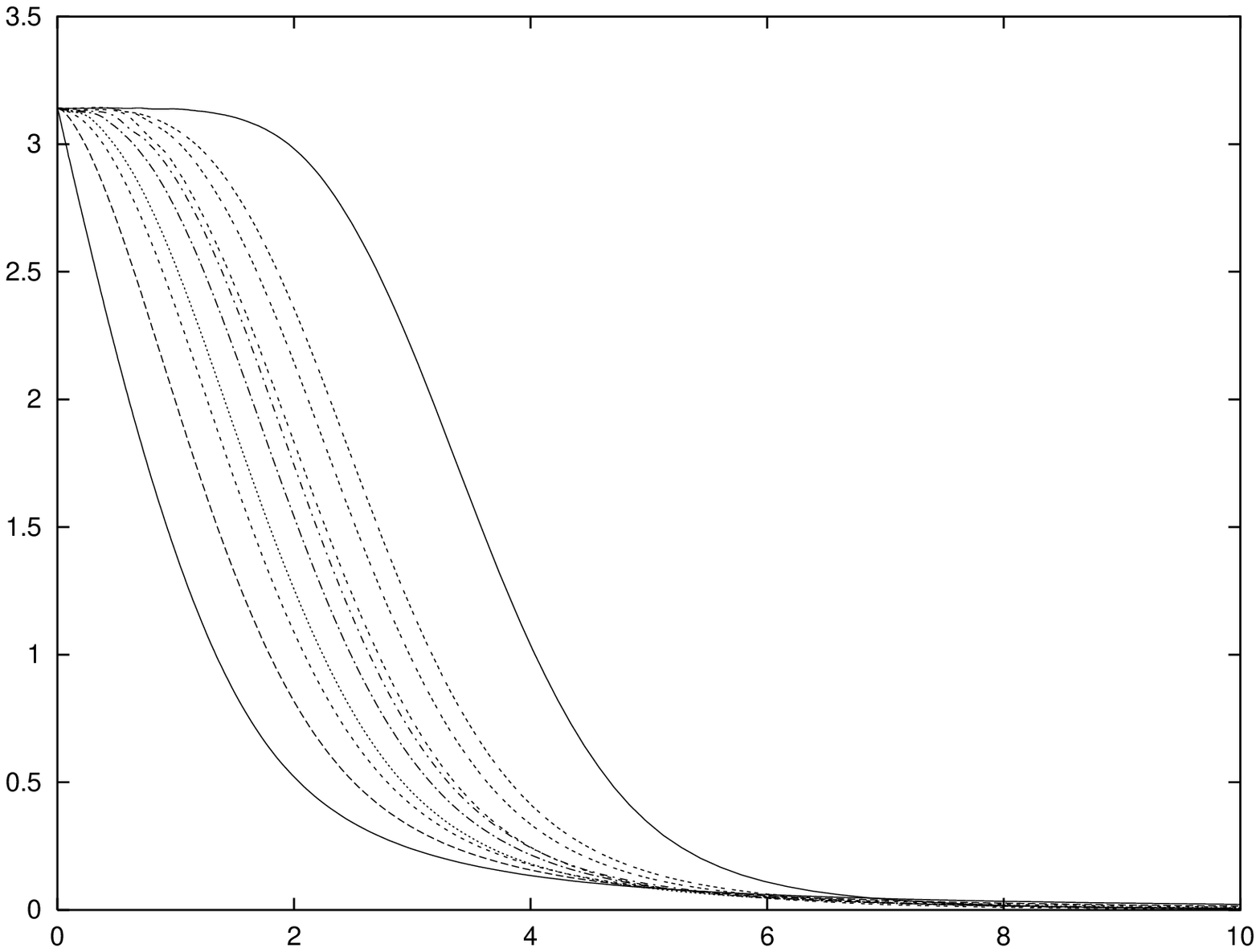,bbllx=89pt,bblly=387pt,bburx=555pt,bbury=744pt,
width=15cm}\\[10pt] 
{\bf Figure 2}
\end{center}
\end{figure}

\begin{figure}[p]
\begin{center}
\epsfig{file=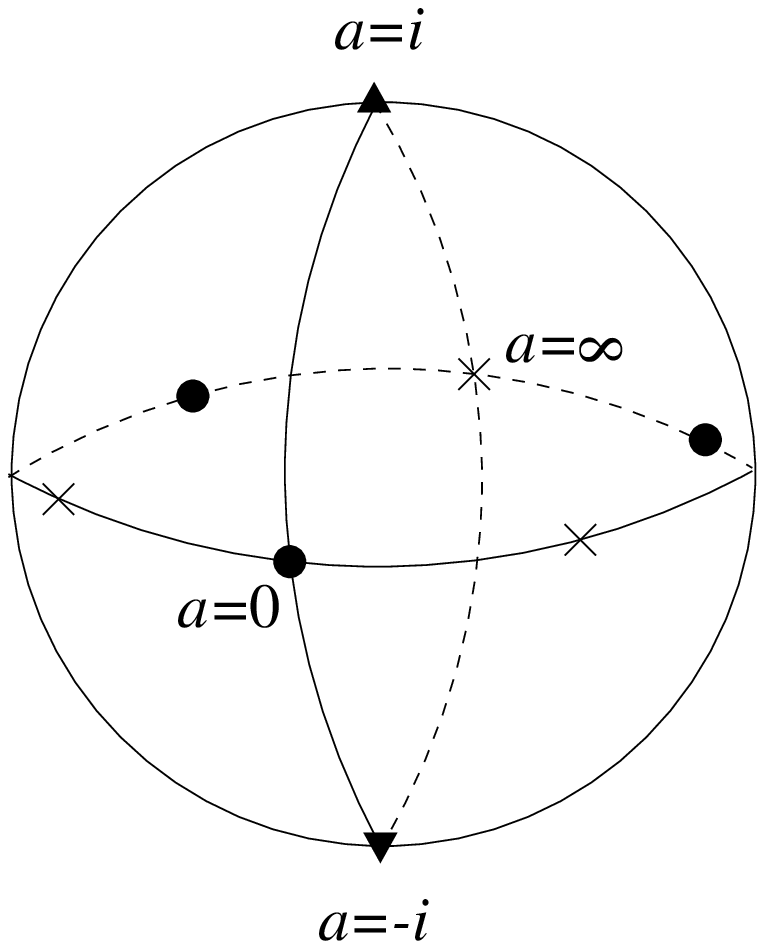,bbllx=198pt,bblly=260pt,bburx=413pt,bbury=499pt,
width=8cm}\\[10pt] 
{\bf Figure 3}
\end{center}
\end{figure}

\begin{figure}[p]
\begin{center}
\epsfig{file=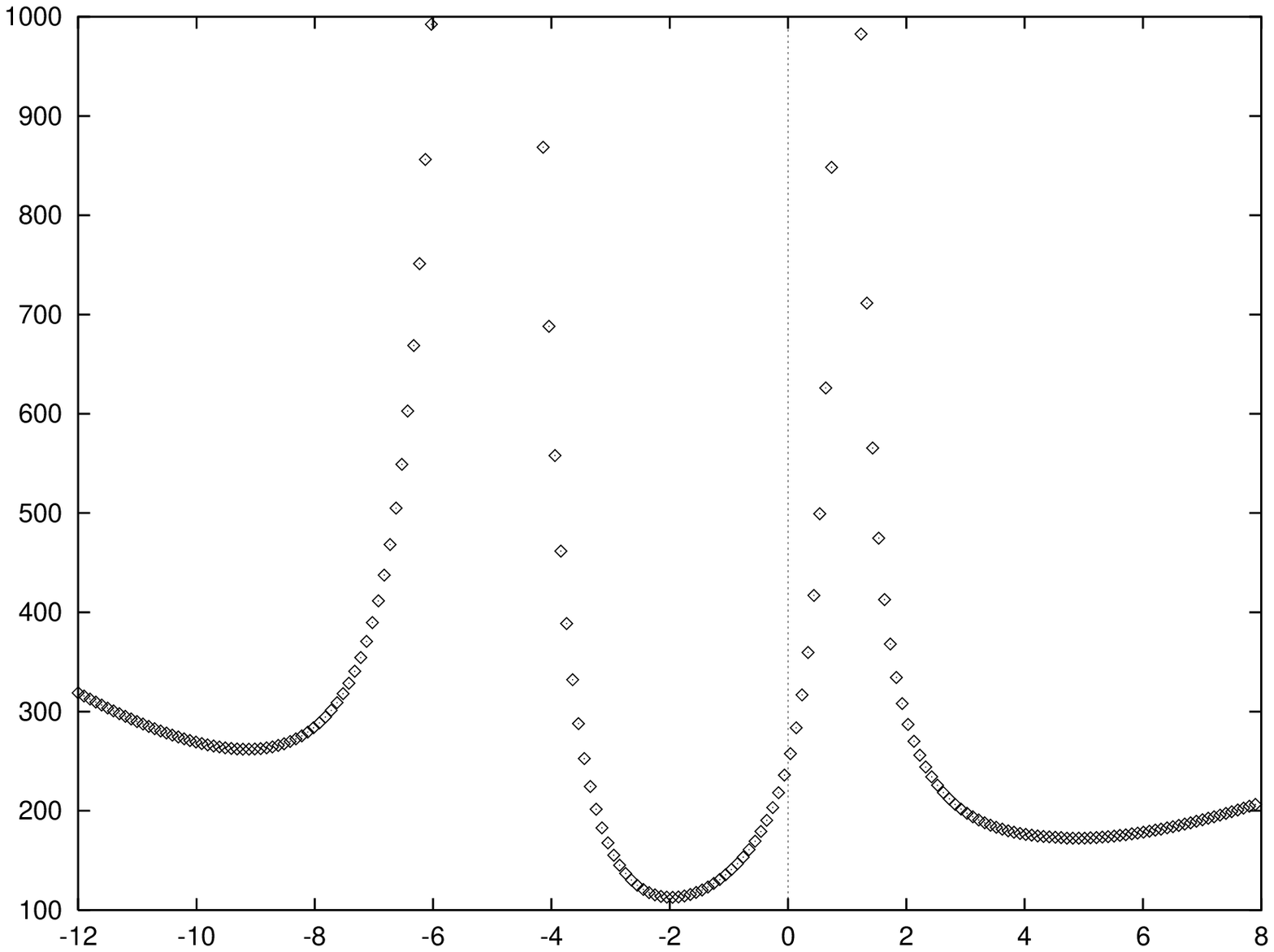,bbllx=90pt,bblly=390pt,bburx=553pt,bbury=746pt,
width=15cm}\\[10pt]
{\bf Figure 4}
\end{center}
\end{figure}

\end{document}